\documentclass[aps,prd,twocolumn,superscriptaddress,floatfix,amsmath,amssymb]{revtex4-1}
\usepackage[english]{babel}
\usepackage{graphicx}
\usepackage{dcolumn}
\usepackage{bm}
\usepackage{verbatim}
\usepackage{mathrsfs}
\usepackage{natbib}
\usepackage{color}
\usepackage{epstopdf}

\newcommand{\integral}[3]{\int_{#2}^{#3} \text{d} #1}

\newcommand{\ket}[1]{\left| {#1} \right\rangle}

\newcommand{\ematriz}[3]{\left\langle {#1} \left|{#2}\right|{#3}\right\rangle}

\def\slashchar#1{\setbox0=\hbox{$#1$} 
\dimen0=\wd0 
\setbox1=\hbox{/} \dimen1=\wd1 
\ifdim\dimen0>\dimen1 
\rlap{\hbox to \dimen0{\hfil/\hfil}} 
#1 
\else 
\rlap{\hbox to \dimen1{\hfil$#1$\hfil}} 
/ 
\fi}

\begin{document}
\title{Unruh-DeWitt detector response along static and circular-geodesic trajectories for Schwarzschild-AdS black holes}
\author{Keith K. Ng}
\affiliation{Dept. Physics and Astronomy, University of Waterloo, Waterloo, ON, N2L 3G1, Canada}
\quad\\
\author{Lee Hodgkinson}
\affiliation{School of Mathematical Sciences, University of Nottingham, Nottingham, NG7 2RD, UK}
\author{Jorma Louko}
\affiliation{School of Mathematical Sciences, University of Nottingham, Nottingham, NG7 2RD, UK}
\author{Robert B. Mann}
\affiliation{Dept. Physics and Astronomy, University of Waterloo, Waterloo, ON, N2L 3G1, Canada}
\affiliation{Institute for Quantum Computing, University of Waterloo, Waterloo, Ontario, N2L 3G1, Canada}
\affiliation{Perimeter Institute for Theoretical Physics, Waterloo, ON, N2L 2Y5, Canada}
\author{Eduardo Mart\'{i}n-Mart\'{i}nez}
\affiliation{Dept. Applied Math., University of Waterloo, Waterloo, ON, N2L 3G1, Canada}
\affiliation{Institute for Quantum Computing, University of Waterloo, Waterloo, Ontario, N2L 3G1, Canada}
\affiliation{Perimeter Institute for Theoretical Physics, Waterloo, ON, N2L 2Y5, Canada}

\begin{abstract}
We present novel methods to numerically address the problem of characterizing the  response of particle detectors in curved spacetimes. These methods allow for the integration of the Wightman function, at least in principle, in rather general backgrounds. In particular we will use this tool to further understand the nature of conformal massless scalar Hawking radiation from a Schwarzschild black hole in anti-de Sitter space. We do that by studying an Unruh-DeWitt detector at rest above the horizon and in circular geodesic orbit. The method allows us to see that the response rate shows peaks at certain characteristic frequencies, which correspond to the quasinormal modes (QNMs) of the spacetime. It is in principle possible to apply these techniques to more complicated and interesting physical scenarios, e.g. geodesic infall or multiple detector entanglement evolution, or the study of the behaviour of quantum correlations in spacetimes with black hole horizons.

\end{abstract}

\maketitle

\section{Introduction}
In recent years, there has been renewed interest in verifying the existence of Hawking radiation in spacetimes with black hole event horizons using the Unruh-DeWitt detector formalism. For instance, a recent  proposal \cite{PhysRevD.89.104002} has allowed for an insightful study of the thermal response of static and circular-geodesic particle detectors in Schwarzschild backgrounds, and there are new and promising results in progress regarding a detector model that is free from infrared divergences \cite{BenitoJorma}, which may be helpful in studying the response of particle detectors across event horizons.

The Hawking effect was first discovered in Schwarzschild spacetime in 1974  \cite{hawking1974}. In essence, Hawking argued that  black holes would radiate as though they had a temperature. Soon afterward, Unruh \cite{unruh1976} suggested the idea of a model particle detector, in order to operationalize the idea of ``observing'' the radiated particles. This method was recently applied in \cite{PhysRevD.89.104002} to a study of the Schwarzschild spacetime.

In this paper, we will analyze the radiation emitted by a black hole in a 4-dimensional asymptotically anti-de Sitter space by means of the vacuum response of a particle detector in this background. This spacetime, often called ``Schwarzschild-Anti-de Sitter'', or SAdS, has been examined by many other authors; however, previous work has mostly focused on other aspects of the spacetime, such as characterizing the decay of scalar modes \cite{chan1996}, analyzing its thermodynamics \cite{hawking1982}, calculating quasinormal frequencies \cite{hh1999,daghigh2008}, and so on. While the Hawking radiation of the SAdS spacetime has previously been studied through other methods, e.g. \cite{hawking1982,hubeny2009}, we believe that our application of particle detectors to the spacetime is novel, and shows new insights.

We model the particle detector with the Unruh-DeWitt model \cite{DeWitt}, which consists of a two-level system with a monopole coupling to a scalar field.  Although simple, this model is known to capture the fundamental features of the light-matter interaction \cite{ScullyBook} when no orbital angular momentum exchange between atomic electrons and the electromagnetic (EM) field is involved \cite{Wavepackets,Alvaro}. The coupling is given by the interaction Hamiltonian
\begin{equation}
H_{\text{int}}(\tau)=c\chi(\tau)\mu(\tau)\phi(x(\tau)),
\end{equation}
where $c$ is a coupling constant, $\chi(\tau)$ is the switching function, $\mu$ is the detector's monopole moment operator, $x(\tau)$ is the trajectory of the detector, and $\tau$ is the proper time of the detector (henceforth simply `proper time'). If the initial state of the joint system is $\ket{\Psi}\otimes\ket{0}_d$ where $\ket{\Psi}$ is a Hadamard state, on which the expectation of the energy-momentum tensor is non-singular, and if $c$ is small, we can write the transition probability (summing over all final field configurations) to first order in perturbation theory as \cite{louko2007}
\begin{equation}
P(E)=c^2 \left\vert_{d}\ematriz{0}{\mu(0)}{1}_d \right\vert^2 F(E),
\end{equation}
where the response function $F(E)$ is independent of the physical details of the detector, aside from its dependence on the detector's energy gap $E$. For instance, this form holds whether the detector is a two-level system or a harmonic oscillator; the differences only become apparent at higher order in perturbation theory and do not yield  a qualitative difference as compared with the two-level quantum emitter \cite{UdWGauss}. For this reason, abusing notation, the response function itself is often simply called the `probability' \cite{louko2007}. The response function can be written as
\begin{align}
F(E)&=\lim_{\epsilon \rightarrow 0} \integral{u} {-\infty}{\infty}\,\chi(u)\nonumber\\ &\times\integral{s}{-\infty}{\infty}\, \chi(u-s)e^{-iE s}W_{\epsilon}(u,u-s)
\end{align}
where $W(u,u')=W(x(u),x(u'))$ is the pullback of the Wightman function $W(x,x')=\ematriz{\Psi}{\phi(x)\phi(x')}{\Psi}$ to the detector trajectory and $\epsilon$ parametrizes its regularization.

In the special cases where the detector is static, or on a circular geodesic orbit, and its response integrated over all times, no special considerations are required to regularize the Wightman function or to control possible divergences related to the switching function; in this case, the Wightman function only depends on the proper time between points, $W(u,u-s)=W(s)$, as discussed in \cite{loukowip}. The response function can then be written as
\begin{equation}
F(E)=\lim_{\epsilon \rightarrow 0} \integral{u} {-\infty}{\infty}\,\integral{s}{-\infty}{\infty}\, e^{-iE s}W_{\epsilon}(s),
\end{equation}
and taking the time derivative (i.e. dropping the $u$ integral) yields the transition rate
\begin{equation}
\label{fdot}
\dot{F}(E)=\lim_{\epsilon \rightarrow 0}\integral{s}{-\infty}{\infty}\, e^{-iE s}W_{\epsilon}(s).
\end{equation}
More precisely, we have taken the limit in which the detector is on for an infinite time; this removes any transient effects due to switching that may induce additional detector excitation \cite{louko2007}. However, there may still be other features inherent in the response function which contain information about the particular spacetime background, as we will see.

Our efforts focus first on finding the solutions to the Klein-Gordon equation in Schwarzschild-anti-de Sitter space, which will allow us to evaluate the Wightman function, and second on using  this to calculate the response of the detector. We then explore some possible interpretations of the numerical results.

\section{Calculating the Transition Rate}
\subsection{Basic equations}

The spacetime known as ``Schwarzschild-AdS'', or simply SAdS, has the following metric:
\begin{equation}
ds^2=-f(r)dt^2+f(r)^{-1}dr^2 + r^2 d\Omega^2_2,
\end{equation}
where the lapse function $f(r)$ is given by
\begin{equation}
f(r)=\frac{r^2}{R^2}+1-\frac{r_0}{r},
\end{equation}
$R=\sqrt{-3/\Lambda}$ is the AdS characteristic length, and $r_0=2M$. Since, without losing generality, we are free to choose an arbitrary value for one of the length scales (equivalent to assuming some system of units), for convenience we set $R=1$.

This spacetime has a few notable features. Like AdS, its `conformal infinity' is timelike; that is, signals can propagate from spatial infinity to any point in Schwarzschild-AdS in finite coordinate time (i.e. finite $t$). Like Schwarzschild, it has an event horizon $r_+$ --- namely, the (real) root of the lapse function $f(r)$. The domain of interest is $r_+<r<\infty$. Note that we can write $r_0=r_+(r_+^2+1)$.

We can also define the tortoise coordinate, defined by $dr^*=dr/f(r)$; this new parameterization reaches a finite value as $r \rightarrow \infty$, and $r^* \rightarrow -\infty$ as $r \rightarrow r_+$. Since $r^*$ is defined up to a constant, we choose to set $r^* = 0$ at infinity; in other words,
\begin{equation}
r^*=-\int^{\infty}_{r}\frac{dr'}{f(r')}.
\end{equation}
In terms of this tortoise coordinate, we can write the (time-independent) one-dimensional Klein-Gordon equation as
\begin{equation}
\label{scatter}
[\partial_{r^*}^2 + \omega^2 -\tilde{V}(r^*)]\tilde{R}=0,
\end{equation}
where the effective potential 
$$\tilde{V}(r^*)=f(r^*)V(r^*)=f(r)\left(\frac{l(l+1)}{r^2}+\frac{r_0}{r^3}\right)$$
vanishes near the horizon.  In the particular case we consider here, that of the massless conformal scalar, the effective potential has a finite value near infinity as well; specifically, $\tilde{V}(r
\rightarrow\infty)=l(l+1).$ This implies that much like the case of the Schwarzschild black hole in flat space, there are modes defined for $r<\infty$ where $\Phi \sim r^{-1}e^{-i\omega (t \pm r^*)}$, which we will call `in' and `out' in analogy.

We also define the null coordinates,  $u = t - r^*$ and $v=t+r^*$, and the Kruskal coordinates 
\begin{align}
U&=-\exp(-2\pi T_H u) \\
V&=\exp(2\pi T_H v)
\end{align}
where
\begin{align}
T_H &= \frac{1}{4\pi}f'(r_+)\nonumber\\
&= \frac{3r_+^2 + 1}{4\pi r_+}
\end{align}
is the usual Hawking temperature.

Most papers analyzing SAdS space have focused on minimally coupled fields; see \citep{berti2009} for a recent review. This type of coupling generalizes quite readily to massive fields and fields of non-zero spin, e.g. gravitational waves. However, in this paper, we will instead focus on the conformally coupled massless scalar field. The conformal coupling was chosen because it most closely mimics the more astrophysically relevant case of the Schwarzschild black hole in flat space---namely, that the effective potential takes a finite value at infinity (if not necessarily zero), and that the effective potential always has a maximum outside the horizon.

Since the conformal infinity of an asymptotically anti-de Sitter space is timelike, we must specify a boundary condition at infinity. We will take the usual Dirichlet boundary conditions---that is, where mode functions vanish at infinity. However, it is easier to start by analyzing the usual incoming and outgoing modes as if conformal infinity were an actual boundary, and then find a linear superposition satisfying the physical boundary conditions.
In these coordinates, if we write out our modes as \citep{hh1999}
\begin{align}
\nonumber w^{in}_{\omega lm}&=(4\pi\omega)^{-1/2}r^{-1}\psi^{in}_{\omega l}(r) Y_{lm}(\theta,\phi)e^{-i\omega (t+r^*)}\\
w^{out}_{\omega lm}&=(4\pi\omega)^{-1/2}r^{-1} \psi^{out}_{\omega l}(r) Y_{lm}(\theta,\phi)e^{-i\omega (t-r^*)},\label{inandout}
\end{align}
our radial equations then become
\begin{align}
f(r)\frac{d^2}{dr^2}\psi^{in}_{\omega l}(r)+[f'(r)-2i\omega]\frac{d}{dr}\psi^{in}_{\omega l}(r)\nonumber\\
-V(r)\psi^{in}_{\omega l}(r) = 0 \\
f(r)\frac{d^2}{dr^2}\psi^{out}_{\omega l}(r)+[f'(r)+2i\omega]\frac{d}{dr}\psi^{out}_{\omega l}(r)\nonumber\\-V(r)\psi^{out}_{\omega l}(r) = 0,
\end{align}
where
\begin{equation}
V(r)=\frac{l(l+1)}{r^2}+\frac{r_0}{r^3}
\end{equation}
for a conformal coupling. Note that $\psi(r)$ approaches a finite value near the horizon; for simplicity we will set $\psi(r_+)=1$.

We make a note here regarding the effective potential, $\tilde{V}(r)=f(r)V(r)$. In contrast to the minimally coupled case, here the effective potential goes to a finite value as $r\rightarrow \infty$. Furthermore, this effective potential always has a maximum above the event horizon for any $r_+$; this is in contrast to the minimally coupled case, where the effective potential for a  sufficiently large black hole will not have a local maximum above the event horizon \cite{berti2009}. Therefore, we do not expect to see phenomena related to the phase transition.

As stated earlier, precisely because we want spatial infinity to be reflecting, these `in' and `out' modes are not typically valid solutions. They do behave correctly everywhere \textit{except} spatial infinity; if we were to apply a conformal transformation to the interval of interest to make it finite, the in and out modes would simply take the wrong value at the point corresponding to spatial infinity. Nevertheless, we can find a linear superposition of them which satisfies the boundary condition. Notice that for real $\omega$, the equations governing the in and out modes are complex conjugates; so, in that case $\psi_{\omega l}^{out}=\psi_{\omega l}^{in*}$.

We now characterize the physical modes $w_{\omega l m}$ in terms of the in and out modes defined in \eqref{inandout}. For $\omega$ positive,  the $\psi$ parts of the in and out modes are complex conjugates. Hence, if we can determine $\theta_0 = \text{ph} [\psi^{in}_{\omega l}(r \rightarrow \infty)]$, then we know that a solution to the Klein-Gordon equation satisfying the boundary conditions is
\begin{align}
\label{inplusout}
w_{\omega lm}&=(4\pi\omega)^{-1/2}r^{-1}e^{-i\omega t}Y_{lm}(\theta,\phi)\nonumber\\
&\times(- i)\left(e^{-i\theta_0}e^{-i\omega r^*}\psi^{in}_{\omega l} - e^{i\theta_0}e^{i\omega r^*}\psi^{out}_{\omega l} \right)
\end{align}
or simply
\begin{align}
w_{\omega lm}&=(4\pi\omega)^{-1/2}r^{-1}e^{-i\omega t}Y_{lm}(\theta,\phi)\nonumber\\
&\times 2\text{Im}\left(e^{-i(\theta_0+\omega r^*)}\psi^{in}_{\omega l}\right).
\end{align}
For brevity, let us define the time-independent radial part of the mode as
\begin{equation}
\label{modeR}
R_{\omega l}(r)=r^{-1} 2\text{Im}\left(e^{-i(\theta_0+\omega r^*)}\psi^{in}_{\omega l}\right).
\end{equation}
The reason we included the $2$ is so that $\tilde{R}_{\omega l}=rR_{\omega l}$ satisfies the Schr\"{o}dinger normalization,
\begin{equation}
\integral{r^*}{-\infty}{0}\tilde{R}_{\omega l}\tilde{R}^*_{\omega' l}=2\pi\delta(\omega-\omega').
\end{equation}
Note that \eqref{modeR} immediately implies that the time-independent radial part of the mode is real.

A cursory examination of the equations governing the metric and the modes suggests that rescaling to a different value of $R$ is fairly simple, with a couple caveats. Time and space coordinates scale in the obvious way: $t \rightarrow \sigma t$, $r \rightarrow \sigma r.$ As one might expect, temperature, energy, and transition rate scale inversely with $\sigma$, i.e. $T_H \rightarrow T_H/\sigma$, $\omega \rightarrow \omega / \sigma$ and $\dot{F}(E) \rightarrow \dot{F}(E/\sigma)/\sigma$. Therefore, $E/T_H$ is invariant under scaling, as is the product of $\dot{F}$ and any of the three lengths $r_+$, $r_0$, and $R$. Notably, $f(r) \rightarrow f(\sigma r)$; the lapse function at equivalent radii is scale-invariant. In this paper, we will use $\omega R$ and $R\dot{F}$ to refer to the dimensionless energy and transition rate respectively.

\subsection{The Hartle-Hawking-like vacuum}

As usual, we can write the field operator as
\begin{equation}
\label{phix}
\phi(x)=\sum_{l=0}^{\infty}\sum_{m=0}^{\infty}\int_{0}^{\infty}\!\!\!d\omega\left(a_{\omega l m}w_{\omega l m} + a_{\omega l m}^{\dagger}w^*_{\omega l m}\right).
\end{equation}
In order to calculate the response function, we then need to calculate how the Wightman function depends on the modes. The derivation works in a similar way as in the asymptotically flat Schwarzschild case;  the details can be found in Appendix \ref{deriv}. The result is that
\begin{align}
W(x,x')&=\sum_{l=0}^{\infty}\sum_{m=-l}^l\int_{0}^{\infty}\frac{d\omega}{2\sinh(\omega/2T_H)}\nonumber\\
&\left[e^{\omega/2T_H}w_{\omega l m}(x)w^*_{\omega l m}(x')\right.\nonumber\\
&\left.+ e^{-\omega/2T_H}w^*_{\omega l m}(x)w_{\omega l m}(x')\right].
\end{align}

We can then write the Wightman function in terms of the radial modes defined in \eqref{modeR}:
\begin{align}
W(x,x')=\sum_{l=0}^{\infty}\sum_{m=-l}^l\int_{0}^{\infty}\frac{d\omega}{8\pi\omega\sinh(\omega / 2T_H)}\nonumber\\
\times\left[e^{\omega/2T_H-i\omega(t-t')}Y_{lm}(\theta,\phi)Y^*_{lm}(\theta',\phi')R_{\omega l}(r) R_{\omega l}(r')\right.\nonumber\\
\left.+e^{-\omega /2T_H +i\omega(t-t')}Y^*_{lm}(\theta,\phi)Y_{lm}(\theta',\phi')R_{\omega l}(r)R_{\omega l}(r')\right].
\end{align}
This expression is almost identical to that of the Schwarzschild case (see \cite{loukowip})---in fact, it \textit{is} identical, after substituting the appropriate $T_H$ and $R_{\omega l}(r)$ functions. Of course, the key difference is that we only have one set of basis functions. Note that this expression for the Wightman function allows us to use essentially the same expression for the transition rate of the static detector as in the Schwarzschild case, with the substitutions noted above.

In the specific case of the static detector, we can simplify even further. The  proper time between $t$ and $t'$ is then just $s=\sqrt{f(r)}(t-t')$, i.e. $(t-t')=s/\sqrt{f(r)}.$ By spherical symmetry, it suffices to consider the case where $\theta=\theta'=0$. In that case, 
 \[Y_{lm}(\theta=0,\phi)=\delta_{m,0}\sqrt{\frac{2l+1}{4\pi}},\]
and thus (the pullback to the worldline of) the Wightman function, $W(s)=W(u,u-s)$, may be written
\begin{align}
W(s)=&\sum_{l=0}^{\infty}\int_{0}^{\infty}\frac{(2l+1)\,d\omega}{32\pi^2\omega\sinh(\omega / 2T_H)}\nonumber\\
&\times\left[e^{\omega/2T_H-i\omega s/\sqrt{f(r)}}\right.\nonumber\\
&\left.+e^{-\omega /2T_H +i\omega s/\sqrt{f(r)}}\right]R^2_{\omega l}(r)\nonumber\\
=&\sum_{l=0}^{\infty}\int_{0}^{\infty}\frac{(2l+1)\,d\omega}{16\pi^2\omega\sinh(\omega / 2T_H)}\nonumber\\
&\times\cos\left[\omega\left(\frac{s}{\sqrt{f(r)}}+\frac{i}{2T_H}\right)\right]R^2_{\omega l}(r)\nonumber\\
=&\sum_{l=0}^{\infty}\int_{0}^{\infty}d\omega\,\frac{(2l+1)}{16\pi^2\omega}R^2_{\omega l}(r)\nonumber\\
&\times\left[\coth(\omega/2T_H)\cos(\omega s/\sqrt{f(r)})\right.\nonumber\\
&\left.-i \sin(\omega s/\sqrt{f(r)})\right]
\end{align}
where $r$ is the radius at which the static detector is located. We can then simply substitute this into \eqref{fdot} to calculate the transition rate; following the derivation in \cite{loukowip}, this yields
\begin{equation}
\label{fdotstat}
\dot{F}(E)=\frac{1}{2E} \frac{1}{e^{E/T_{loc}}-1}\sum_{l=0}^{\infty}\frac{2l+1}{4\pi}R^2_{\tilde{\omega}l}(r),
\end{equation}
where $\tilde{\omega}=\sqrt{f(r)}E$ and $T_{loc}=T_H/\sqrt{f(r)}$. Notice that since the remaining integral in \eqref{fdot} evaluates to a Dirac delta, we only need to evaluate the mode at one value of $\omega$ for each $l$.

We can also consider the case where the detector is in a circular geodesic orbit at radius $r$. For convenience, we will write:
\begin{align}
\label{circab}
a&:=dt/d\tau=\sqrt{\frac{2r}{2r-3r_0}}\nonumber\\
b&:=d\phi/d\tau=\sqrt{\frac{r_0+2r^3}{r^2(2r-3r_0)}}
\end{align}
The transition rate of a detector in a circular geodesic orbit can then be found to be \cite{loukowip}
\begin{align}
\dot{F}(E)=&\sum_{l=0}^{\infty}\sum_{m=-l}^{l}\frac{(l-m)!}{(l+m)!}\frac{2l+1}{16\pi}|P_l^m(0)|^2
 \nonumber \\
 &\times \left(\Theta(\omega_-)\frac{e^{2\pi \omega_-}}{a\omega_-\sinh(2\pi\omega_-)}R^2_{\omega_- l}(r)\right.
 \nonumber\\
 &+\left.\Theta(\omega_+)\frac{e^{-2\pi \omega_+}}{a\omega_+\sinh(2\pi\omega_+)}R^2_{\omega_+ l}(r)\right)
\end{align}
where $\omega_{\pm} = \frac{mb\pm E}{a}$ is a function of $m$, and $P_l^m(x)$ is the associated Legendre polynomial. Notice that, since we must sum over a number of $m$ proportional to $l$ for each $l$, the total number of modes evaluated is of order $l^2$. This is in contrast to the static case, where we only needed to calculate one mode for every $l$, namely at $m=0$. (While we can take advantage of certain symmetries of $P_l^m(0)$ to shorten the calculation, the general scaling relation still holds.)

At this point, it should be noted that $V_l(r)$ near the horizon, $r\approx r_+$, behaves like
\begin{equation}
V_l(r)\approx \frac{1}{r^2}\left(l(l+1)+(r_+^2+1)\right).
\end{equation}
This is rather problematic: It means that for large $r_+$---the case which is most interesting from the AdS/CFT perspective---we will need to calculate the Wightman function and modes to high angular momentum, of order $r_+$. In particular, since the circular geodesic calculation requires $O(l^2)$ mode calculations, that calculation can quickly become intractable. However, this is unavoidable given the mode separation method. On the other hand, for small $r_+$, we will only need to worry about very small angular momenta.

\subsection{The Boulware vacuum}

We may also consider the vacuum in which static observers outside the horizon observe no particles; this is known as the Boulware vacuum. The Wightman function of the Boulware vacuum is
\begin{align}
W(x,x')&=\left\langle\Psi\right|\phi(x)\phi(x')\left|\Psi\right\rangle \nonumber\\
&=\sum_{l=0}^{\infty}\sum_{m=-l}^l\int_{0}^{\infty}\frac{d\omega}{4\pi\omega}\nonumber\\
&\times\left[e^{-i\omega(t-t')}Y_{lm}(\theta,\phi)Y^*_{lm}(\theta',\phi')R_{\omega l}(r) R_{\omega l}(r')\right],
\end{align}
while the static detector response rate is simply
\begin{equation}
\dot{F}(E)=\Theta(-E)\frac{1}{2|E|} \sum_{l=0}^{\infty}\frac{2l+1}{4\pi}R^2_{\tilde{\omega}l}(r),
\end{equation}
and the circular detector response is (using the quantities defined in \eqref{circab})
\begin{align}
\dot{F}(E)=&\frac{1}{a}\sum_{l=0}^{\infty}\sum_{m=-l}^{l}\frac{(l-m)!}{(l+m)!}\frac{2l+1}{8\pi\omega_-}|P_l^m(0)|^2 \nonumber\\
&\times\Theta(\omega_-)R^2_{\omega_- l}(r)
\end{align}
where $\omega_{-} = \frac{mb- E}{a}$ as before.

Comparing the Boulware and Hartle-Hawking vacua allows us to determine the effect of Hawking radiation, separate from any other possible effects on the response of the detector.

\subsection{ Numerical methods}

 Two independent numerical methods were used in order to calculate the transition functions of the detector. One was a close adaptation of the methods used in \cite{PhysRevD.89.104002, loukowip}; briefly, we evaluated from initial conditions via series solutions for the in and out modes \eqref{inandout} up from the horizon, and the physical modes \eqref{modeR} down from infinity, then found the phase $\theta_0$ which related the two modes via Wronskian methods. The other is described in detail in this section. It is similar to a previous method first used by Horowitz and Hubeny \cite{hh1999} to find the quasinormal modes of a minimally coupled scalar in Schwarzschild-AdS; we used it to find modes of the conformally coupled scalar.

We apply a transformation to the radial part of the solutions to the Klein-Gordon equation, substituting $x=1/r$. Letting $x_+=1/r_+$, we find that the in and out radial solutions satisfy
\begin{align}
\label{adsx}
s(x)\frac{d^2}{dx^2}\psi^{in}_{\omega l}(x)+\frac{t^{in}(x)}{x-x_+}\frac{d}{dx}\psi^{in}_{\omega l}(x)\nonumber\\
 + \frac{u(x)}{(x-x_+)^2}\psi^{in}_{\omega l}(x) = 0\\
s(x)\frac{d^2}{dx^2}\psi^{out}_{\omega l}(x)+\frac{t^{out}(x)}{x-x_+}\frac{d}{dx}\psi^{out}_{\omega l}(x)\nonumber\\
 + \frac{u(x)}{(x-x_+)^2}\psi^{out}_{\omega l}(x) = 0
\end{align}
where
\begin{align}
s(x)&=\frac{x_+^2+1}{x_+^3}x^4+\frac{1}{x_+^2}x^3+\frac{1}{x_+}x^2\\
t^{in}(x)&=3r_0 x^4 -2x^3-2x^2i\omega\\
t^{out}(x)&=3r_0 x^4 -2x^3+2x^2i\omega\\
u(x)&=(x-x_+)V(x).
\end{align}

We then expand the solutions around the horizon $x_+$:
\begin{align}
\label{adsinexp}
\psi^{in}_{\omega l}(x)&=\sum_{n=0}^{\infty}a_n^{in}(x-x_+)^n\\
\label{adsoutexp}
\psi^{out}_{\omega l}(x)&=\sum_{n=0}^{\infty}a_n^{out}(x-x_+)^n;
\end{align}
the coefficients then are governed by the recurrence relations
\begin{align}
\label{ain}
a_{n}^{in}=-\frac{1}{P^{in}_n}\sum_{k=0}^{n-1}[k(k-1)s_{n-k}+kt^{in}_{n-k}+u_{n-k}]a^{in}_k\\
\label{aout}
a_{n}^{out}=-\frac{1}{P^{out}_n}\sum_{k=0}^{n-1}[k(k-1)s_{n-k}+kt^{out}_{n-k}+u_{n-k}]a^{out}_k
\end{align}
where
\begin{align}
s(x)&=\sum_{n=0}^{4}s_n(x-x_+)^n\\
t^{in}(x)&=\sum_{n=0}^{4}t^{in}_n(x-x_+)^n\\
t^{out}(x)&=\sum_{n=0}^{4}t^{out}_n(x-x_+)^n\\
u(x)&=\sum_{n=0}^{4}u_n(x-x_+)^n\\
P^{in}_n&=n(n-1)s_0+nt^{in}_0\\
P^{out}_n&=n(n-1)s_0+nt^{out}_0.
\end{align}
Note that the recurrence relations \eqref{ain}, \eqref{aout} only involve a finite number of $a_k$ terms (five, in this case), as $s(x)$, $t(x)$, and $u(x)$ are all quartic polynomials.

As mentioned earlier, these solutions do not satisfy the boundary condition at infinity. Specifically, because of the structure of \eqref{adsx}, the expressions in \eqref{adsinexp}, \eqref{adsoutexp} will diverge at $x=0$. However, summing over a finite number of terms $N$ at $x=0$ allows us to find a linear combination of those modes which vanishes at $x=0$; in other words, we can solve for $A$, $B$ such that
\begin{equation}
A\sum_{n=0}^N [a^{in}_n (-x_+)^n] + B\sum_{n=0}^N [a^{out}_n (-x_+)^n] =0.
\end{equation}
We can then increase $N$ and verify that the linear combination still vanishes. In the particular case where $\omega$ is real, the in and out modes are complex conjugates, so this must be possible; we use the approach described in the previous section, culminating in \eqref{modeR}. 

For smaller values of $\omega$, smaller $l$, and near the horizon, the power series expression for $\psi^{in}_{\omega l}$ can be found with a reasonable value of $N$. However, for larger values of $\omega$, larger $l$, and for larger radii, convergence takes a very large number of terms; in particular, it is more efficient to compute the values of the modes at spatial infinity using another approach, e.g. using the power series expansion at finite distance and numerically integrating the differential equation to spatial infinity.

Note that this method is somewhat different from the approach taken by Horowitz and Hubeny \cite{hh1999}. Since they were interested in quasinormal modes, they only considered modes that were ingoing at infinity; they then solved for complex $\omega$ such that $\sum_{n=0}^N[a^{in}_n (-x_+)^n]=0$. Our $\omega$, on the other hand, can take any  real value, and we allow for superpositions of in and out modes.

\section{Numerical Results}

\begin{figure}
	\includegraphics[scale=0.55]{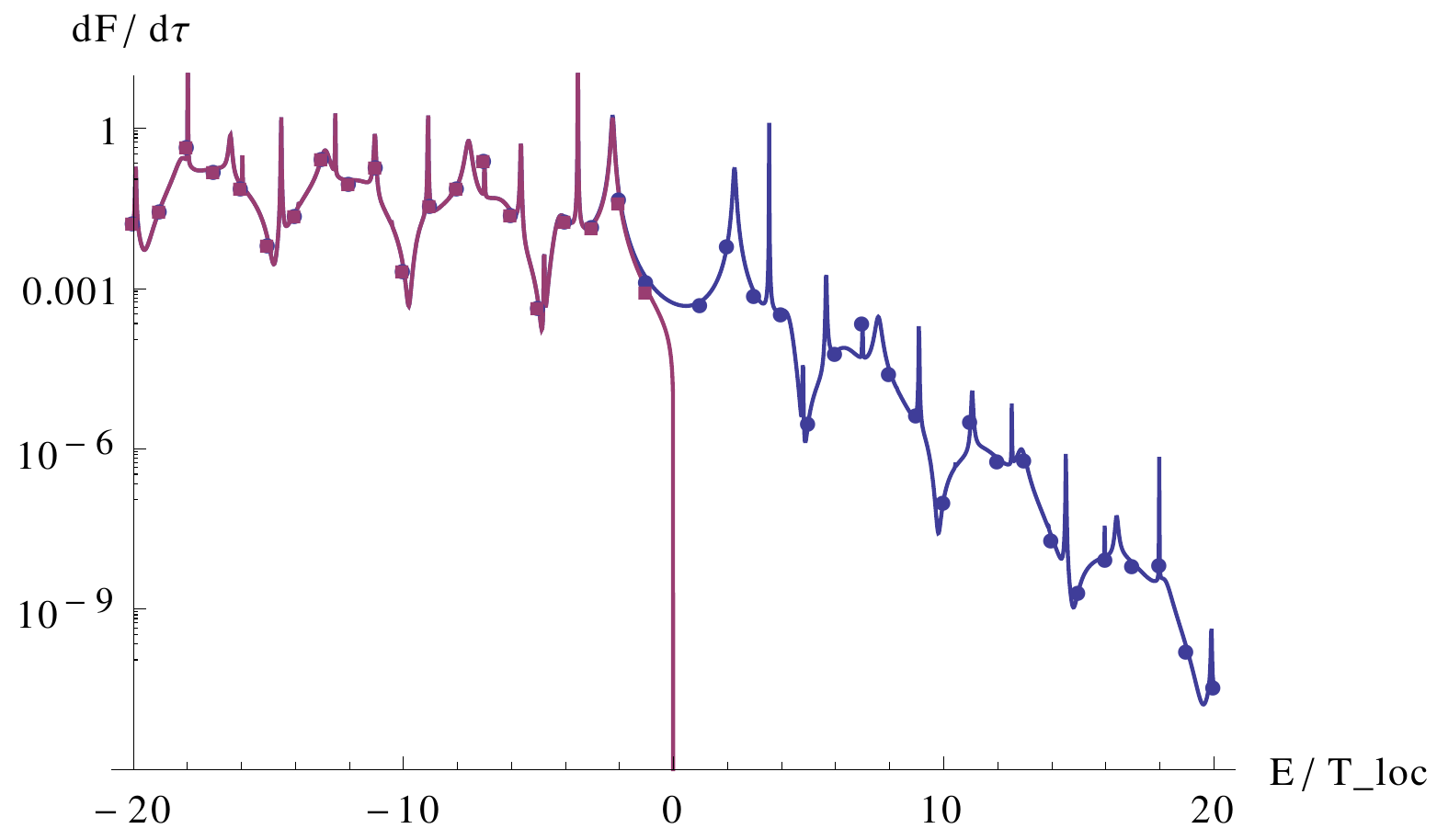}
	\caption[total transition $r_+=0.1$]{ The total static transition rate for $r_+=0.1, r=1$. Hartle-Hawking vacuum in blue circles, Boulware in red squares. Note Boulware transition is zero for positive $E$.}
	\label{total0.1}
\end{figure}

Using the previously outlined methods, we numerically calculated the response rate of the static detector for various values of the relevant parameters.  The Wronskian method performed better for small $r_+/R$, while the Horowitz-Hubeny method was faster for larger $r_+/R$. Both methods were in agreement over the range of parameters where they could both be applied. As mentioned earlier, the appropriate expression is \eqref{fdotstat}. First, plotted in Fig. \ref{total0.1} is the transition rate for $r_+=0.1$, $r=1$, summing from $l=0$ to 4. The horizontal axis indicates the relative detector energy gap $E/T_{loc}$ and the vertical axis indicates $R\dot{F}(E)$. While there are a number of different scales in this problem, assuming units such that $R=1$ is the simplest way to make the transition rate dimensionless. The blue curve marked with circles indicates the Hartle-Hawking vacuum response, while the red curve with squares indicates the Boulware vacuum response. Note that for this and the following graphs, the markers are intended as an aid to identifying the curves; the actual density of data points is much higher.
Convergence of the $l$ sum can be easily verified by carrying out the summation to higher $l$ order and comparing.

There are a couple of notable features in the transition rate. First, note that for very large negative energy gap, corresponding to an initially excited detector, the Boulware and Hartle-Hawking vacuum responses are almost identical. Second, a number of spikes are observed, both in the Boulware and Hartle-Hawking vacuum response; in fact, the responses of both vacua are quite similar in the region of the plot where spikes occur. There do not appear to be any other interesting features in the regime where the different vacua produce different results: therefore we will henceforth focus on the Hartle-Hawking response.

\begin{figure}
	\includegraphics[scale=0.55]{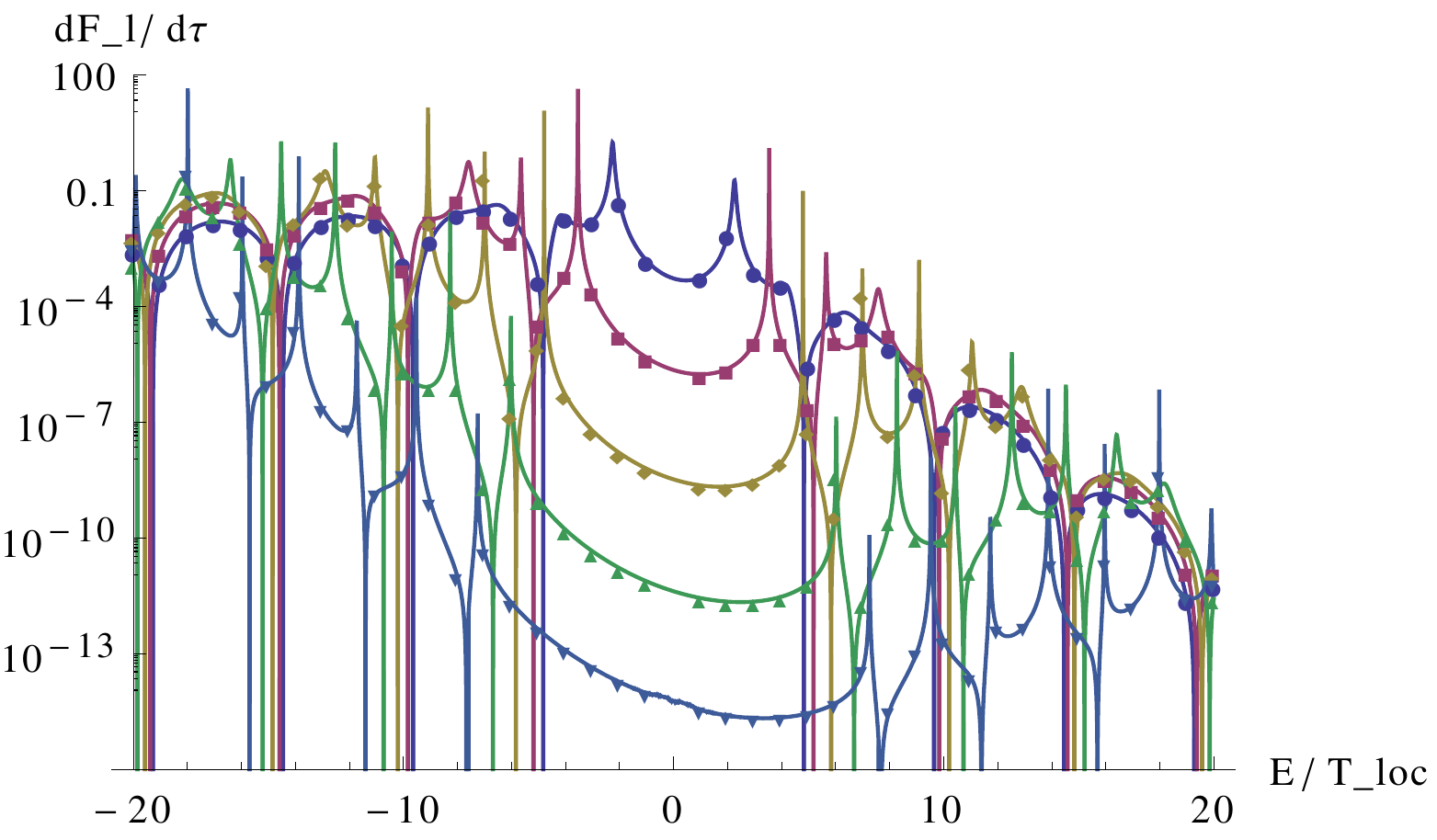} 
	\caption{ Static transition rate contributions for $r_+=0.1, r=1, l=0,1,..,4$ from top to bottom.}
	\label{ells0.1}
\end{figure}

Next, there are also a number of `dips' in the response. This has a simple explanation: since the modes are real, there must be some energy for which a zero of a mode crosses the location of the detector. Therefore, if we plotted the contributions of different $l$ individually, we would see the transition rate go to zero. Indeed, this is visible in Fig. \ref{ells0.1}, where $l=0$ is the top blue line with circles, $l=1$ is the next line down in red with squares, and so on to $l=4$.

\begin{figure}
	\includegraphics[scale=0.55]{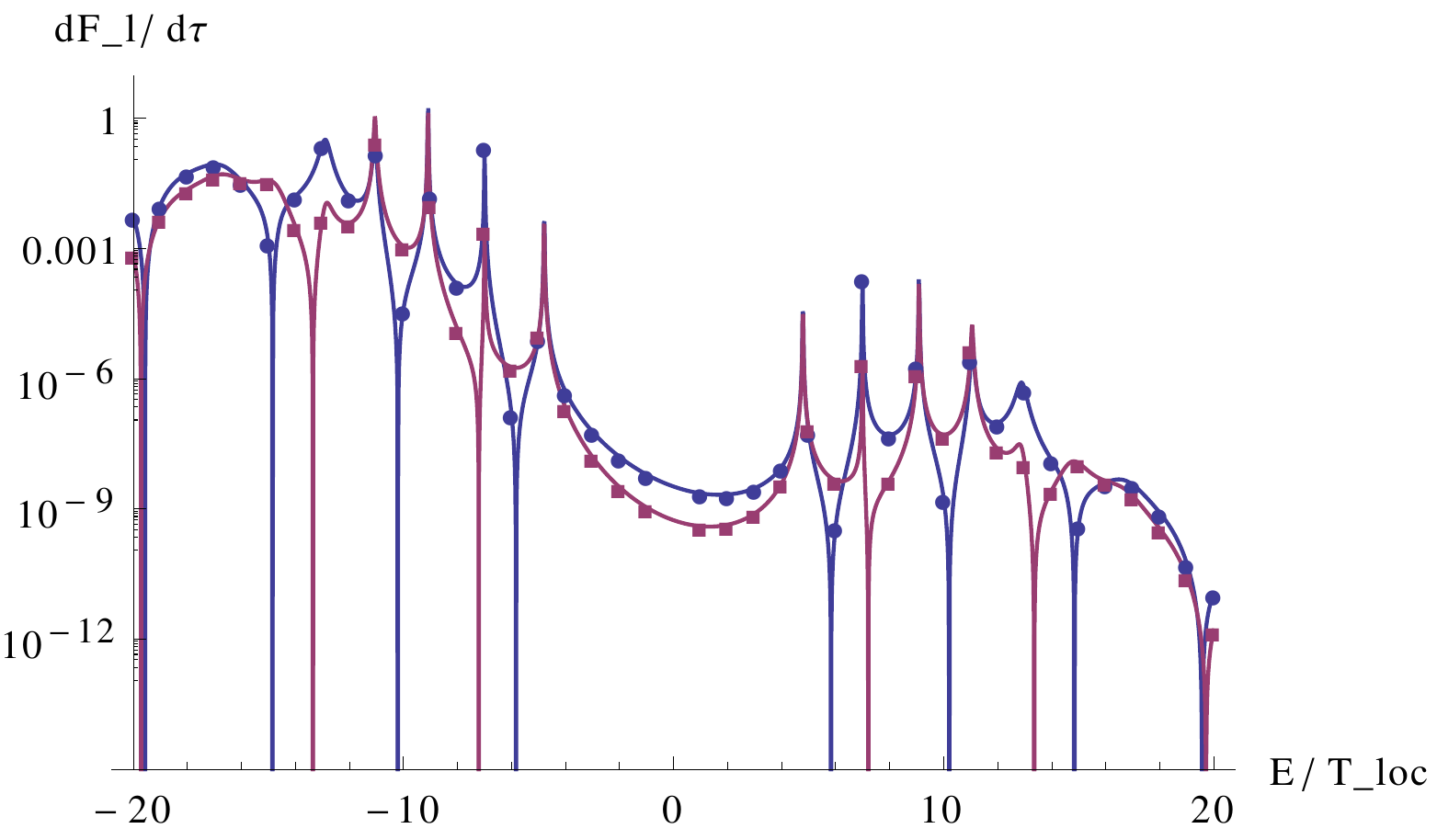} 
	\caption{. Transition rate for different $r$. The static transition rate contribution for $r_+=0.1, l=2, r=1$ in blue circles, $r=1.5$ in red squares. }
	\label{rs0.1}
\end{figure}

Recall that the transition rate for detector gap $E$ only depends on the modes at a particular energy $\tilde{\omega}=(E/T_{loc})T_H$. Therefore, the best way to compare transition rates at two different radii is to plot both of them against $E/T_{loc}$ as we have done in Fig. \ref{rs0.1}, for $l=2$; $r=1$ is in blue circles, $r=1.5$ in red squares. Notice how the peaks occur at the same locations, even as the zeroes shift. This suggests that our explanation for the zeroes is correct; the peaks will be addressed later.

\begin{figure}
	\includegraphics[scale=0.55]{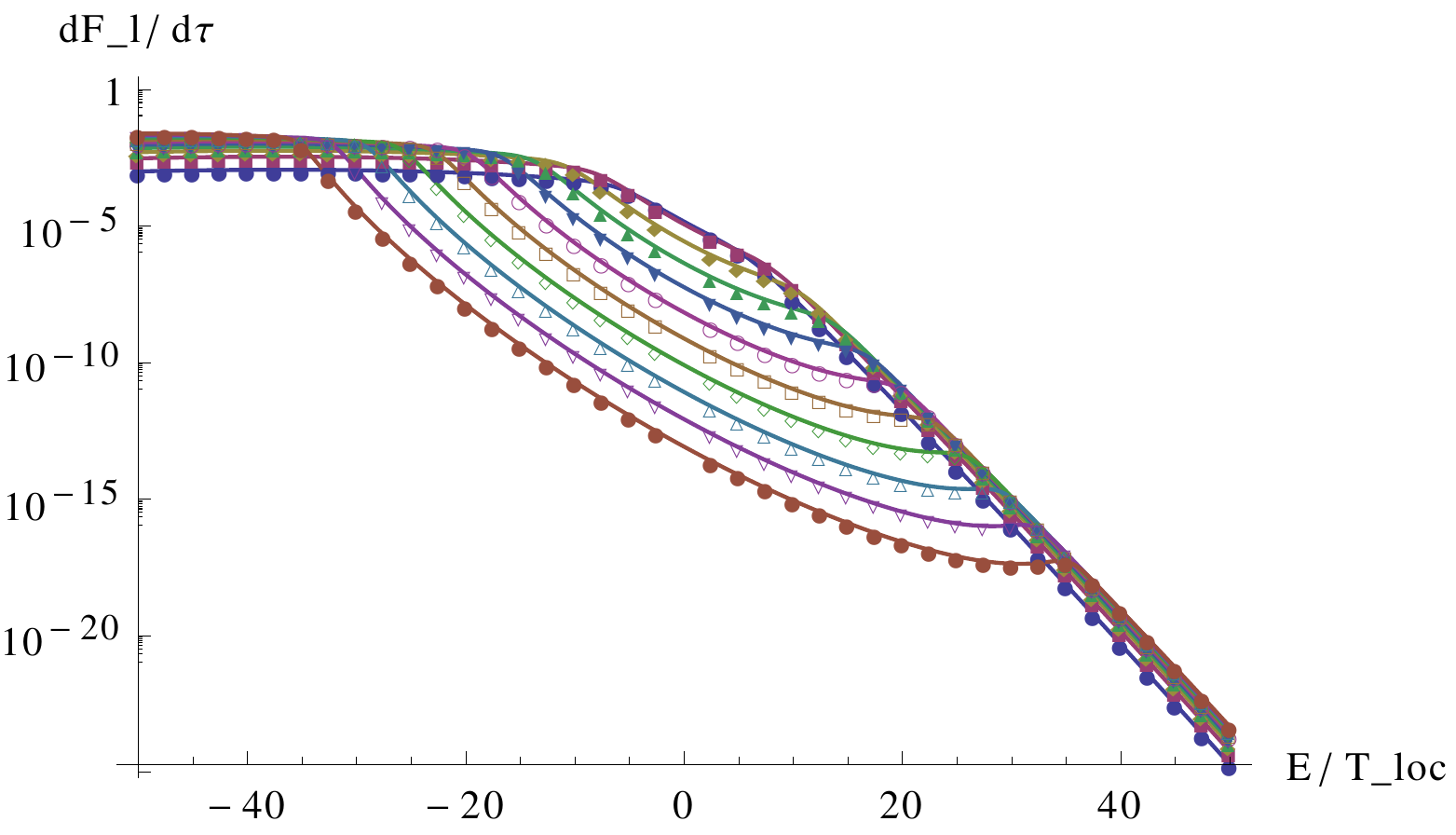} 
	\caption{ : Static transition rate contributions for $r_+=1$, $r=10$, $l=0,1,...,10$ from top to bottom.}
	\label{total1}
\end{figure}

Next, we plot the contributions of $l=0,1,2,...,10$ for $r_+=1$, $r=10$ in Fig. \ref{total1}. Note that $r/r_+=10$ is kept constant; using this scaling allows us to compare situations with different black hole sizes, without worrying about scaling the detector into the horizon. In Fig. \ref{total1} once again $l=0$ is the top line, $l=1$ is the next line down, and so on.

Notice that the larger black hole appears to allow contributions from higher $l$ modes. This makes sense, since we noted that near the horizon (and thus near the peak of the effective potential), the dependence of the potential on $l$ decreases as $r_+$ increases. At low energies, as we would expect, the contribution from $l=0$ is largest, followed by $l=1$, and so forth. However, at higher energies, it appears as though the contributions for various $l$ are comparable, up to some maximum. Of course, this means that if we wish to calculate the total transition rate for $E/T_{loc}>30$, we will need to consider higher $l$ modes.  This phenomenon is also visible in Fig. \ref{ells0.1}, but to a lesser extent; while the presence of the peaks confuses things somewhat, the $l=2$ contribution does start being smaller than the $l=0$ contribution at low energy, becoming comparable at larger energy.

\begin{figure}
	\includegraphics[scale=0.55]{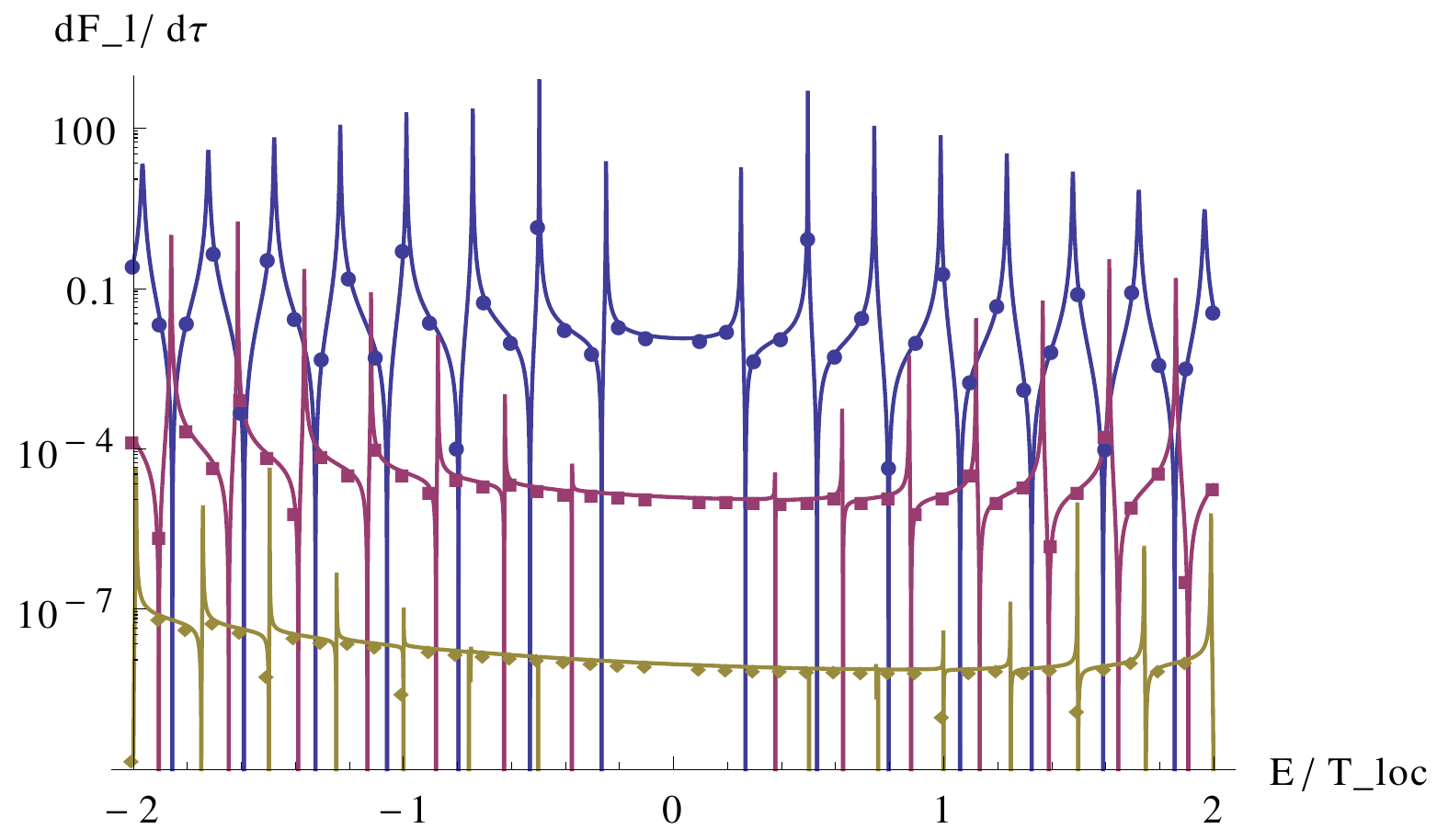}
	\caption[l transition $r_+=0.01$]{The static transition rate contributions for $r_+=0.01, r=0.1, l=0,1,2$ from top to bottom.}
	\label{ells0.01}
\end{figure}

The next graph, Fig. \ref{ells0.01}, shows the contributions of $l=0,1,2$ for $r_+=0.01$, $r=0.1$, with $l=0$ at the top in blue circles, $l=1$ below in red squares, and $l=2$ at the bottom in yellow diamonds. First, the suppression of higher $l$ modes at smaller $r_+$ is clearly visible. Second, the `spikiness' of the graph appears to have increased from the $r_+=0.1$ case---not only are the peaks at low energy sharper, but the peaks appear to be present at higher energies than in the $r_+=0.1$ case. 

 However, the graph is in some ways misleading. The higher $l$ modes are much, much spikier than the lower $l$ modes: the barely visible spike at $l=2$, $E/T_{loc}\sim 1/2$ actually goes up to almost $10^9$, although we require more than ten digits of precision in $E/T_{loc}$ to find the maximum of the peak properly. Unfortunately using this level of precision for the graph is not feasible, so the maximum heights of the peaks shown in the graphs are not completely accurate. The situation is comparable to that of Fig \ref{ells0.1}, in which higher $l$ modes can dominate at the peaks, but the peaks themselves are far thinner here.

In Fig. \ref{ells0.01} our choice of scale for the horizontal axis appears to have placed the peaks in approximately the same places as in the $r_+=0.1$ case, Fig. \ref{ells0.1}; on the other hand, the exponential decrease of the transition rate with respect to increasing $E$ appears to be more gradual. In other words, the relationship between the scale of the exponential decrease and the scale of the peaks changes as we manipulate the ratio of the black hole size to the AdS length $(r_+/R)$. Going in the opposite direction, for $r_+=1$, na\"{i}vely applying scaling suggests that a peak should appear at about $E/T_{loc}=20$; no peak is present, suggesting that the exponential decrease has overwhelmed the peaks entirely.

\begin{figure}
	\includegraphics[scale=0.9]{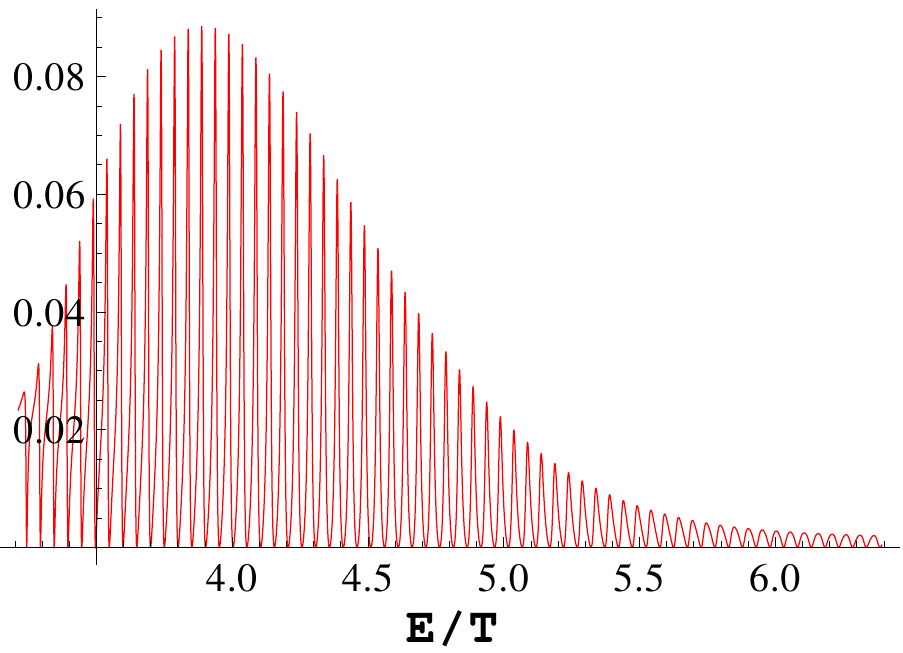}
	\caption[l=0 transition $r_+=0.002$]{The static transition rate contribution of $l=0$ for $r_+=0.002, r=0.02$ at higher energy.}
	\label{leegraph0}
\end{figure}

 The next graph, Fig. \ref{leegraph0}, shows the contributions of $l=0$ for $r_+=0.002$, $r=0.02$, at a higher energy. It demonstrates the persistence of the peaks at higher energies for smaller black holes, as we previously asserted---although the peaks have broadened significantly. The variation in the height of the spikes is actually an interaction between the troughs and peaks - their spacings are slightly different, so the transition rate is smaller in the region where the troughs and peaks line up.

\begin{figure}
	\includegraphics[scale=0.9]{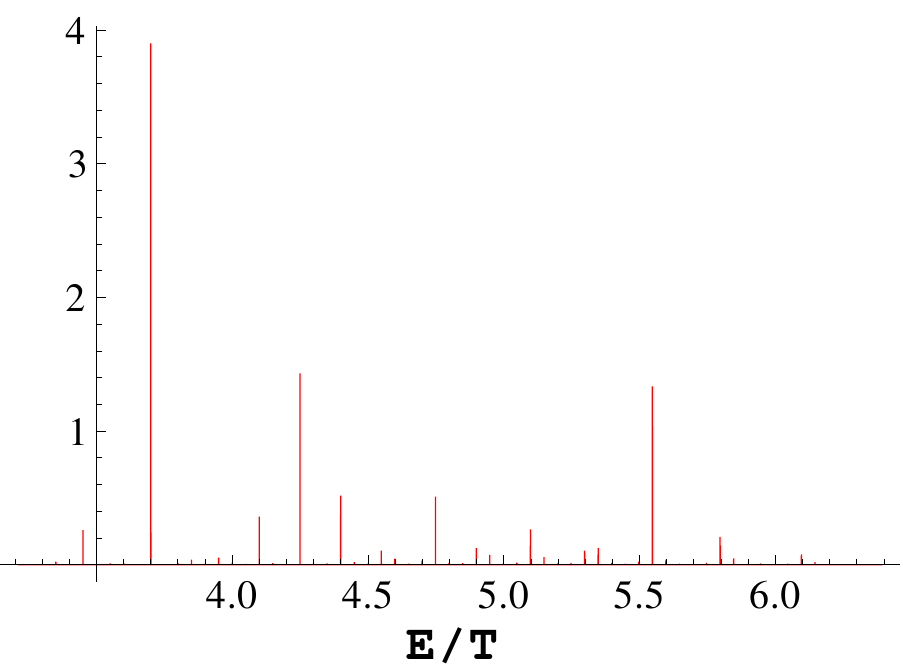}
	\caption[l=2 transition $r_+=0.002$]{The static transition rate contribution of $l=2$ for $r_+=0.002, r=0.02$ at higher energy.}
	\label{leegraph2}
\end{figure}

In Fig. \ref{leegraph2}, the contribution of $l=2$ for $r_+=0.002,$ $r=0.02$ is plotted at very high relative energy. However, despite the high energies involved, it is quite clear that the sharpness and height of the peaks is still very significant. As expected, the peaks are sharper than in the $l=0$ case, Fig. \ref{leegraph0}. Also notable is that once again, the absolute height of the peaks is well above that of the peaks in Fig. \ref{leegraph0}; any computation at this energy will require consideration of $l=2$, and likely much higher $l$ as well.

The observations in the previous paragraphs can be summarized as follows: In general, it appears that as $|E|$ is increased, the spikes visibly become  shorter and broader, to the point where the peaks are not apparent at all; when $r_+$ is decreased,  and as $l$ increases, the spikes become  taller and thinner, and persist at higher $|E|$. This is probably due to a competition between the exponential trend of the transition rate and the sharpness of the peaks: if $r_+$ is large enough, the exponential trend dominates, and the peaks cannot be seen. The \textit{location} of the peaks appears to be on a different energy scale from the exponential decay; the relationship between these two scales changes as we change $r_+/R$.

Our previous observations also seem to suggest a necessary precaution: when $r_+/R$ is very small, the peaks at high $l$ become quite extreme. Therefore, in order to properly represent the sum over all $l$ at some energy, it appears that we must sum over all $l$ which have peaks at lower energies, since we saw earlier that, under certain circumstances, high $l$ mode peaks can dominate over lower $l$ modes.

\begin{figure}
	\includegraphics[scale=0.55]{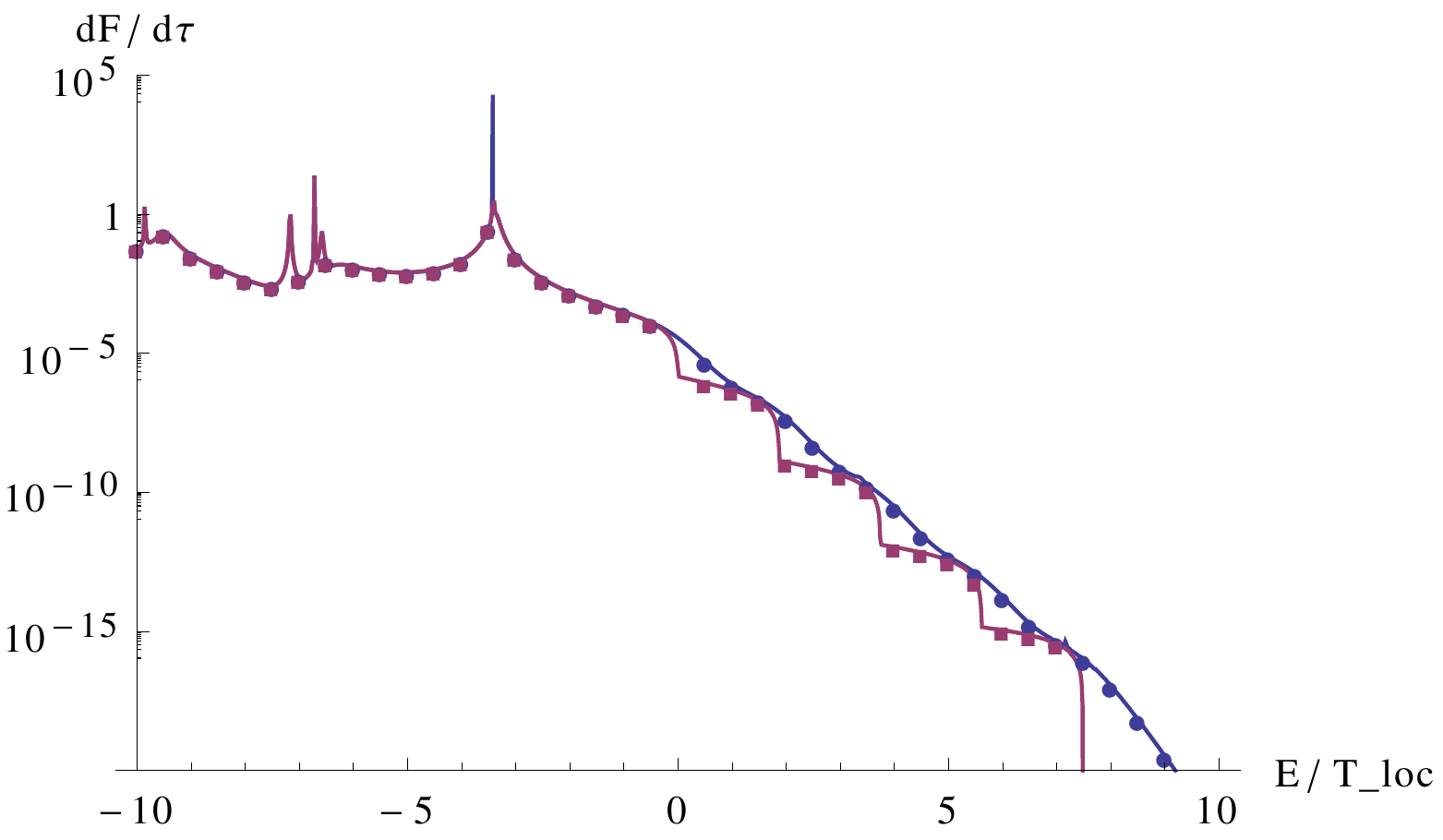}
	\caption{ Total circular geodesic transition rate contributions for $r_+=0.1, r=1.$ Hartle-Hawking vacuum in blue circles, Boulware vacuum in red squares }
	\label{totalcirc}
\end{figure}

We also briefly discuss the circular geodesic case. The transition rate is illustrated in Fig. \ref{totalcirc} for $r_+=0.1$, $r=1$, summing to $l=4$, with the Boulware vacuum in red, and the Hartle-Hawking vacuum in blue. Once again, for almost all $E<0$, the two are almost exactly the same.   Unlike the static case, however, the Boulware transition is nonzero for some $E>0$. We will discuss it more later. 

\begin{figure}
	\includegraphics[scale=0.55]{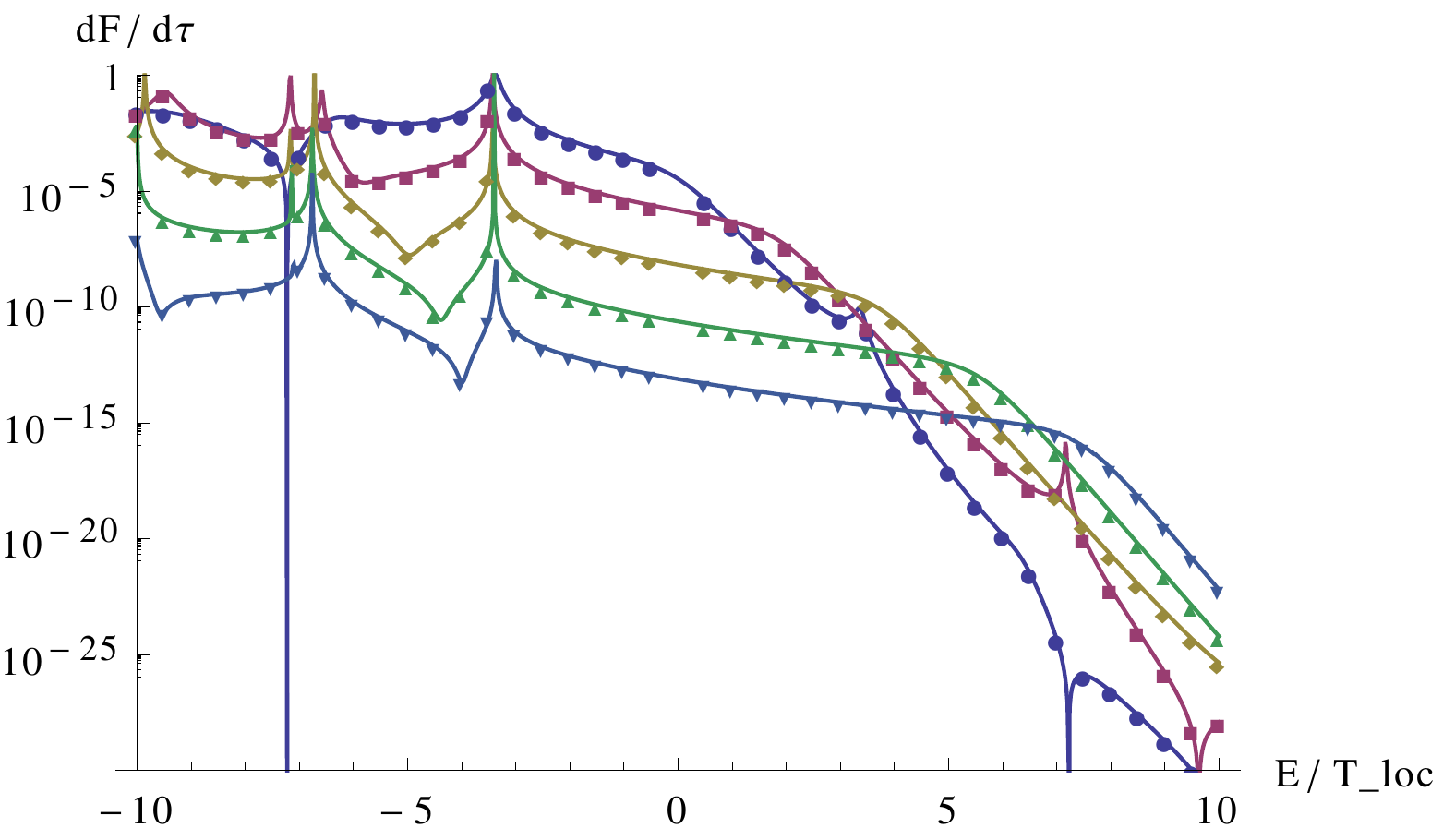}
	\caption{ Hartle-Hawking circular geodesic transition rate contributions for $r_+=0.1, r=1, l=0,1,..,4$ from top to bottom.}
	\label{ellscirc}
\end{figure}

One notable feature is that the Hartle-Hawking transition appears ``shifted'' rightwards compared to the static graphs; it is no longer symmetric about $E=0$, but instead about some positive energy. This is expected, since the detector is accelerating relative to static observers. Plotting the component $l$ individually in Fig. \ref{ellscirc} helps clarify what is going on here; once again, we have $l=0$ in blue circles on the top, $l=1$ in red squares below, and so on to $l=4$.

At this point, it is apparent that each $l$ appears more shifted than the last; the ``centre'' of each curve lies slightly farther to the right as $l$ increases. The shift is such that each $l$ dominates for a particular range of energies. The explanation lies in the definition of $\omega_-=(mb-E)/a$: for any positive $E$ there will be some $m$ such that $\omega_-$ takes its smallest positive value, and thus dominates; but $l \ge |m|$, and so this particular $m$ can only be achieved for sufficiently large $l$.  This is also why the Boulware transition rate appears to go to zero at some finite $E$: If we continued summation to higher $l$, we would see the transition rate stay nonzero at higher $E$. However, since the $\omega_-$ term occurs in both the Boulware and Hartle-Hawking responses, the nonzero transition rate at positive $E$ should not be interpreted as a feature of the Hawking radiation; rather, it is a result of the circular {\it motion}. 

It is also now clear why no peaks were visible for $E>0$. While a small peak is visible in the $l=0$ transition rate, for instance, the strong exponential decay suppresses it. In fact, the suppression is strong enough that the peak is `hidden' by the higher $l$ modes; at the energy where the $l=0$ peak is located, both the $l=1$ and $l=2$ contributions are greater in magnitude, and their exponential trend masks the peak further.

\begin{figure}
	\includegraphics[scale=0.55]{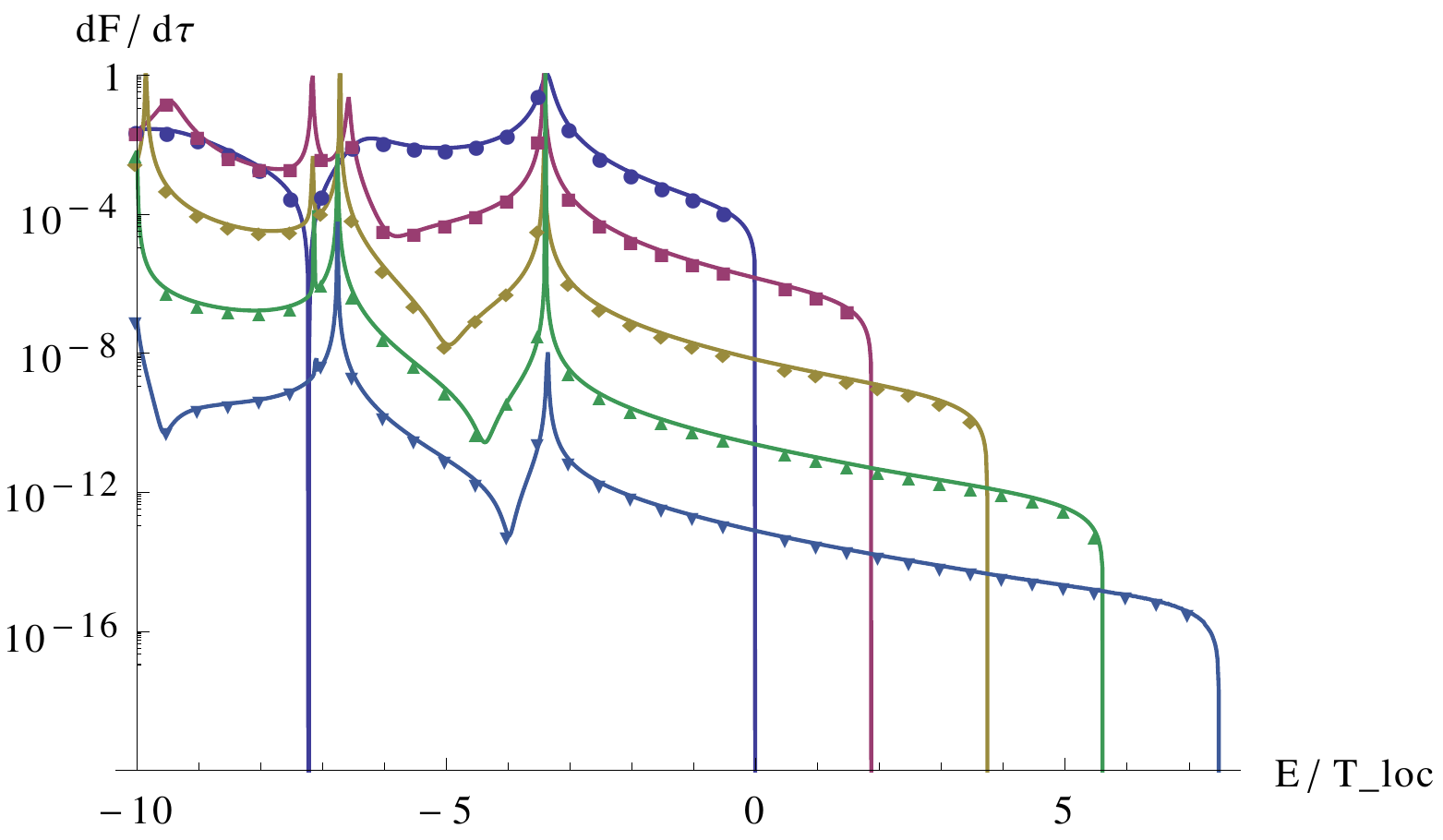}
	\caption{ Boulware circular geodesic transition rate contributions for $r_+=0.1, r=1, l=0,1,..,4$ from top to bottom.}
	\label{ellscircboul}
\end{figure}

For comparison, we include Fig. \ref{ellscircboul} for the Boulware transition rate. Note that the `step' in the transition rate at zero seen in Fig. \ref{totalcirc} is due solely to the $l=0$ contribution; the higher $l$ modes only vanish at higher energy. Specifically, each $l$ contribution vanishes for $E=lb$, since this is the energy such that the highest $\omega_{-}$ becomes zero. This also explains the ``steps'' visible in the positive energy transition rate; each step simply corresponds to an $l$.

Besides that, however, the graphs appear to show precisely the same features as observed in the static detector scenario, e.g. the characteristic peaks. Therefore, we will focus our analysis on the previous case.

\section{Analysis}

We now discuss the peaks present in the static detector transition rate. At an abstract level, the spikes are reflective of resonances of the Klein-Gordon field over this spacetime---in other words, spikes occur when the frequency approaches a quasinormal frequency. While the literature on the analysis of quasi-normal modes (QNMs) is rich, the particular case of conformally coupled scalar fields on SAdS has not been fully explored at the present time---specifically, a table of quasinormal frequencies has not yet been calculated for this particular case. Therefore, we will use an alternate analysis.

Since the Klein-Gordon equations are equivalent to a scattering problem in one dimension, in tortoise coordinates, it makes sense to consider what happens if we have an incident wave from infinity with coefficient $1$---that is, consider the following (approximate) solution to \eqref{scatter}:
\begin{equation}
\tilde{R}^{scatter}_{\omega l}=
\begin{cases} 
\frac{1}{\sqrt{\omega'}}\left(e^{-i \omega' r^*} + Ae^{i \omega' r^*}\right) &r \rightarrow \infty \\
\frac{1}{\sqrt{\omega}}Be^{-i \omega r^*} &r \rightarrow r_+
\end{cases}
\end{equation}
where $\omega'^2 = \omega^2 - l(l+1)$ is the squared wavenumber ``at infinity'', since $\tilde{V}$ does not vanish at infinity. We assume $\omega'^2$ is positive, i.e. $\omega^2 > l(l+1)$. (While the regime  $\omega^2 < l(l+1)$ may \textit{a priori} bear some interesting phenomena, empirically this does not appear to be the case: there is no particular structure at energies below that of the first peak.) This solution to the Klein-Gordon equation is simply the incoming mode $\tilde{R}^{in}_{\omega l}$ from before, up to a constant coefficient. 

There is a subtlety, however: since $r^*$ is finite when $r$ is infinite, the approximation requires that $\omega'$ is sufficiently large that the potential does not change much over a wavelength. Again, in practice this assumption generally appears to be justified. More precisely, we are using a Wentzel-Kramers-Brillouin (WKB) approximation; for wavenumber $k(r^*)=\sqrt{\omega^2-\tilde{V}(r)}$, the validity condition is that $|k'(r^*)|/k^2(r^*)\ll 1/2\pi$, which is valid for small black holes and far from the black hole. For instance, as $r^*\rightarrow 0$, $|k'(r^*)|/k^2(r^*)\rightarrow r_0/2\omega'^{3}.$

Then, since $\psi(r_+) =  1$, we must have
\begin{equation}
\tilde{R}^{in}_{\omega l}=\frac{\sqrt{\omega}}{B}\tilde{R}^{scatter}_{\omega l},
\end{equation}
which in turn implies that
\begin{equation}
\tilde{R}^{in}_{\omega l}=
\begin{cases}
\sqrt{\frac{\omega}{\omega'}}\frac{1}{B}\left(e^{-i \omega' r^*} + Ae^{i \omega' r^*} \right)&r \rightarrow \infty \\
e^{-i \omega r^*} &r \rightarrow r_+
\end{cases}
\end{equation}

Now, recall how we used this to get a mode satisfying the boundary condition: we set $\tilde{R}_{\omega l}=2\,\text{Im}[e^{-i\theta_0}\tilde{R}^{in}_{\omega l}]$. In this case, we can see that $\theta_0 = \text{Arg}[\frac{A+1}{B}]$. So,
\begin{equation}
\tilde{R}_{\omega l}=
\begin{cases}
\sqrt{\frac{\omega}{\omega'}}\frac{2}{|B|}\text{Im}\left(\frac{|A+1|}{A+1}\left(e^{-i\omega' r^*} + Ae^{i\omega' r^*}\right)\right)
&r \rightarrow \infty\\
2 \,\text{Im}[e^{-i (\omega r^* + \theta_0)}] &r \rightarrow r_+
\end{cases}
\end{equation}

We can clarify the situation by rewriting in terms of trigonometric functions. Specifically, near infinity, the physical modes must look like
\begin{align}
\tilde{R}_{\omega l} &\rightarrow \sqrt{\frac{\omega}{\omega'}}\frac{2}{|B|} \text{Im}\left(\frac{|A+1|}{A+1}\left(\frac{A+1}{2}\cos(\omega' r^*)\right.\right.\nonumber\\
& \left.\left.+ i\frac{A-1}{2}\sin(\omega' r^*)\right)\right)\nonumber\\
&= \sqrt{\frac{\omega}{\omega'}}\frac{|A+1|}{|B|}\text{Im}\left(\cos(\omega' r^*) + i\frac{A-1}{A+1}\sin(\omega' r^*)\right)\nonumber\\
&= \sqrt{\frac{\omega}{\omega'}}\frac{|A+1|}{|B|} \text{Re}\left(\frac{A-1}{A+1}\sin(\omega' r^*)\right)
\end{align}
The meaning of the last line is clarified if we use the identity $\frac{|A+1|}{A+1}=\frac{(A+1)^*}{|A+1|}$, which results in
\begin{align}
\label{peakiness}
\tilde{R}_{\omega l} &\rightarrow\sqrt{\frac{\omega}{\omega'}}\frac{1}{|B||A+1|} \text{Re}\left((A-1)(A+1)^*\sin(\omega' r^*)\right)\nonumber\\
&= \sqrt{\frac{\omega}{\omega'}}\frac{1}{|B||A+1|} \text{Re}\left((AA^* + A - A^* - 1)\sin(\omega' r^*)\right)\nonumber\\
&= \sqrt{\frac{\omega}{\omega'}}\frac{|A|^2 - 1}{|B||A+1|}\sin(\omega' r^*)\nonumber\\
&= -\sqrt{\frac{\omega}{\omega'}}\frac{|B|}{|A+1|}\sin(\omega' r^*)
\end{align}
where the last equation follows from the fact that $|A|^2 + |B|^2 = 1$ in  our approximation.
(Of course, near the horizon, $\tilde{R}_{\omega l}\rightarrow -2 \sin(\omega r^* + \theta_0)$, which we previously demanded in order to satisfy normalization.)

An explanation for the peaks now presents itself: we must experience a peak when the reflection coefficient $A$ approaches $-1$, i.e. the phase of $A$ approaches $\pi$. This corresponds to having $\tilde{R}^{in}_{\omega l}(r\rightarrow \infty)$ approach $0$ --- in other words, it is much like having the incoming (at the horizon) mode satisfy the boundary condition at infinity. Of course, those boundary conditions are precisely those satisfied by the quasinormal modes, so we have come full circle.

The previous derivation has a small caveat: we relied on the WKB approximation to determine the behaviour of the mode near infinity. However, the validity condition typically is not satisfied at the particular $r$ of the detector and the energies of the peaks shown in the graphs. We can relax the validity condition by allowing the amplitude to change with $r^*$, which corresponds to taking a higher order WKB approximation; in that case, we would see that the amplitude smoothly interpolates from $2$ near the horizon to a large value at infinity, so a large amplitude at infinity indicates a large amplitude at any intermediate distance, and thus a peak in the detector transition rate.

Besides avoiding the invocation of QNMs, the analysis above additionally allows us to make qualitative predictions regarding the peaks. For instance, assuming the phase of $A$ changes much faster than its magnitude (which appears to be the case when $r_+$ is sufficiently small), the local maxima of the coefficient $C={|B|}/{|A+1|}$ in \eqref{peakiness} occur when $A$ is negative real, and are
 \[C=\frac{|B|}{1-|A|}=\frac{\sqrt{1-|A|^2}}{1-|A|},\]
 while the local minima occur when $A$ is positive real, and are
 \[C=\frac{|B|}{1+|A|}=\frac{\sqrt{1-|A|^2}}{1+|A|}.\]

 The maxima and minima are both $1$ when $|A| = 0$; as $|A| \rightarrow 1$, the maxima monotonically increase towards infinity, while the minima monotonically approach zero. Since we expect $|A| \rightarrow 0$ as $\omega^2 \rightarrow \infty$, peakiness decreases as energy increases; conversely, as energy decreases, peakiness must increase. Of course, there is a peak of lowest energy, i.e. a lowest energy QNM, so there will not be an infinite sequence of higher and higher peaks as $\omega^2 \rightarrow 0$. Additionally, the fact that $l$ corresponds to a higher effective potential suggests that as $l$ increases, the real part of the frequency of the lowest-lying QNM will also increase---in other words, it suggests that the peaks will occur at higher energies. The exact relationship between $A$ and $\omega$, however, must be calculated.

As an aside, if we translate the above predictions into the language of QNMs, we are essentially predicting that as the real part of the QNM increases in magnitude, the imaginary part also increases in magnitude; and that as $l$ increases, so does the real part of the QNM. This agrees with the behaviour of QNMs in SAdS for other couplings (e.g. minimal) noted in the literature (see \cite{berti2009} for a thorough survey).

We may also compare the peaks found here to the more familiar case of normal modes in AdS. Following the approach found in \cite{burgess}, using effective mass $\mu^2=-2R^2$ to yield a conformal coupling, we find that the normal mode frequencies corresponding to our coupling and boundary conditions are 
\begin{equation}
\omega R=2+l+2n.
\end{equation}
Notably, this is quite similar to the frequencies of the minimally coupled modes, $\omega R=3+l+2n$.

In order to translate the peak detector energies $E$ into mode energies, recall that $\tilde{\omega} = \sqrt{f(r)}E$; therefore, 
\begin{equation}
E/T_{loc} = \tilde{\omega}/T_H
\end{equation}
Using this equation, we can observe from the graphs that the peaks converge to the AdS normal conformal modes as $r_+ \rightarrow 0$: for instance, when $r_+=0.01$, $T_H=7.96$, so the first peak ($l=0,n=0$) corresponds to a mode frequency of about $2.0$. This makes sense---the smaller the black hole is, the smaller its ``influence'' over the volume of AdS.

\begin{figure}
\includegraphics[scale=0.5]{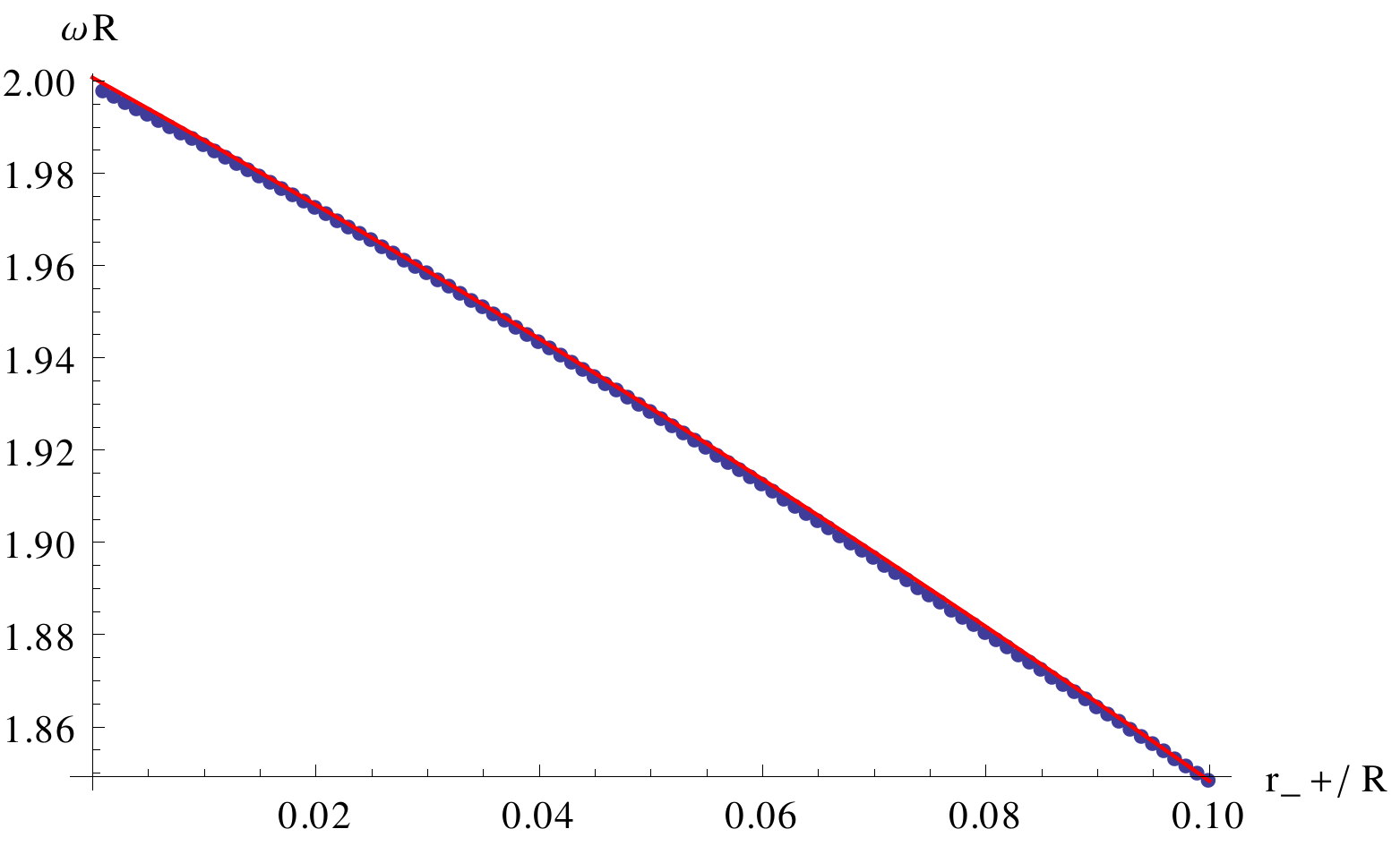}
\caption{  Frequency corresponding to the peak of the static transition rate with respect to black hole size, for small black holes (thick dotted line). A quadratic fit (red solid line) has been overlaid on the peak frequencies.}
\label{peakstat}
\end{figure}

In order to verify the convergence to the AdS normal mode limit, we plotted the location (i.e. the corresponding mode frequency) of the first peak in the $l=0$ transition rate, corresponding to $n=0$, with respect to the black hole size. The results are plotted in Fig. \ref{peakstat}: similarly to the minimally coupled case mentioned in \cite{berti2009}, the trend is linear as $r_+ \rightarrow 0$, with a very slight next-order (i.e. quadratic) term visible; a quadratic fit has been plotted on the graph. This strongly suggests that the frequency of the SAdS QNMs is tied to the AdS scale, rather than the Schwarzschild scale, at least for small black holes.

\begin{figure}
\includegraphics[scale=0.5]{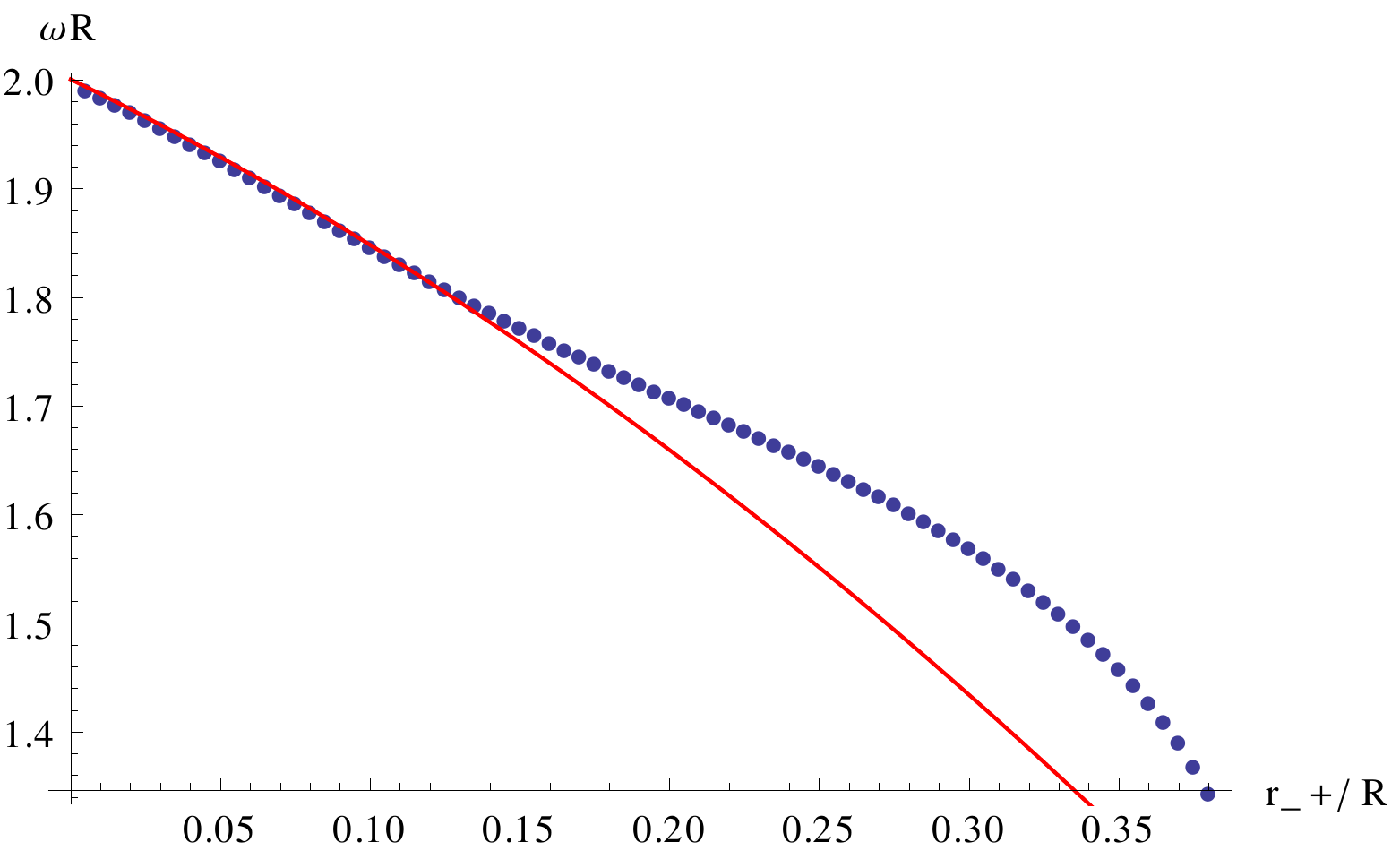}
\caption{  Frequency corresponding to the peak of the static transition rate with respect to black hole size, for larger black holes (thick dotted line). The quadratic fit displayed in Fig. \ref{peakstat} (red solid line) is also plotted here.}
\label{peakstatlarger}
\end{figure}

However, the analogue of the limit $r_+ \rightarrow \infty$ is rather less clear. First, while not shown in the graph in Fig. \ref{peakstat}, when $r_+$ is sufficiently large, the peak disappears; it is simply suppressed by the larger-scale trend of exponential decay in the transition rate. Even before then, some behaviour is visible that departs from the quadratic fit done on the previous graph, as we can see in Fig. \ref{peakstatlarger}; there is an increase in peak frequency over the general trend above the quadratic fit done in the previous case. Also, the peak in the transition rate disappears completely just beyond the end of the region plotted. It is likely that the reason for the deflection is that the peak is being ``shifted'' by the exponential term in the transition rate; in that case, the true location of the QNM no longer corresponds to the peak of the transition rate.

Physically, as $r_+$ is increased, the horizon moves `towards' the SAdS conformal boundary; or, if we choose to keep the horizon radius constant and scale the AdS radius instead, the cosmological constant becomes larger and larger in magnitude, and the conformal boundary moves towards the horizon. This limit is quite unlike the Schwarzschild black hole in flat space. It is not simply a matter of placing a reflecting sphere very near the Schwarzschild black hole in flat space, either---since the effective potential always has a peak within SAdS, any reflecting sphere would have to stay outside the peak of the effective potential in the Schwarzschild-flat case. In fact, this limit is \textit{also} unlike the minimally coupled case: in that case, a sufficiently large SAdS black hole has no local maximum in the effective potential outside the horizon (see e.g. \cite{berti2009}), while our conformally coupled case always does. In the end, it is probably better to consider the limit of a large black hole as a different physical situation entirely; the QNMs in that case may not converge to any more familiar form, and in any case may not be of any relevance to the transition rate.

\section{Conclusion  and Outlook}

We have computed, for the four-dimensional Schwarzschild anti-de Sitter spacetime, the response of an Unruh-DeWitt detector in static and circular geodesic trajectories to a conformally coupled scalar field. The response function bears some sharp peaks with respect to the detector energy gap; we have demonstrated that these spikes are due to quasinormal mode resonances. There are also some troughs in the graph; when the contributions are separated by $l$, it becomes clear that this corresponds to when a zero of the mode function crosses the location of the detector.

We have also attempted to characterize the dependence of the location of the peaks on the radius of the black hole in AdS space. Qualitatively, the spikes are only visible when the black hole is much smaller than the AdS length; as the black hole's size is decreased, the spikes appear sharper and higher. One might have expected a transition between small, intermediate, and large black holes, in analogy with the minimally coupled case. However, this type of transition cannot occur in this case: the effective potential of the conformally coupled scalar field always has a maximum at finite distance, and as a result, no phenomena related to the phase transition are apparent in the conformally coupled scalar field. Our computation of the peak frequencies at various black hole sizes confirms the convergence of the peak frequency as $r_+ \rightarrow 0$ to the AdS normal frequencies; the disappearance of peaks at higher black hole size appears to be mainly due to the dominance of the exponential decay term over the peak, rather than any sort of phase transition of the spacetime as a whole.

We would like to note that the calculation of the static and circular geodesic transition rates is a first step towards characterizing the response of the detector to Hawking radiation on more general trajectories, e.g. radial geodesic infall.  Remarkably, the Unruh-Dewitt detector formalism used here can also be applied to even more general physical scenarios, such as calculating the evolution of the entanglement of two detectors above the black hole, a calculation that would be relevant for the study of the dynamics of correlations and information near black hole horizons. The usage of  these methods in these scenarios may shed light on some of the central mysteries of the black hole, e.g. the question of what happens to information lost beyond the horizon.

\section*{Acknowledgements}
This work was supported in part by the Natural Sciences and Engineering Research Council of Canada.
 L.H. was supported by EPSRC
through a Ph.D. Plus Fellowship at the University of Nottingham.
J.L. was supported in part by STFC (Theory Consolidated Grant
ST/J000388/1). E.M-M. acknowledges the support of the Banting
Postdoctoral Fellowship Programme. Parts of the numerical work were
carried out on the Nottingham High Performance Computing Facility.

\appendix
\section{Derivation of the Hartle-Hawking Wightman Function}
\label{deriv}

The Hartle-Hawking vacuum can be constructed by computing the Bogoliubov coefficients of the external modes with respect to the Kruskal modes \cite{kv2000}. While the Dirichlet boundary condition presents a slight complication, the derivation proceeds much like the Schwarzschild case.

In this particular case, we have what is essentially a reflecting boundary at infinity, and so we only have one basis of modes on each exterior, rather than Schwarzschild's two; the situation is analogous to that of the black hole in a reflecting boundary analyzed in \cite{frolovnovikov}. Consider the mode given in \eqref{inplusout}:
\begin{align}
w_{\omega lm}&=(4\pi\omega)^{-1/2}r^{-1}e^{-i\omega t}Y_{lm}(\theta,\phi)\nonumber\\
&(-i)\left(e^{-i\theta_0}e^{-i\omega r^*}\psi^{in}_{\omega l} - e^{i\theta_0}e^{i\omega r^*}\psi^{out}_{\omega l} \right).
\end{align}
It fulfils the boundary condition at infinity, and is a positive frequency superposition of in and out modes. We can express it in terms of $u,v$ as
\begin{align}
w_{\omega lm}&=(4\pi\omega)^{-1/2}r^{-1}Y_{lm}(\theta,\phi)\nonumber\\
&(-i)\left(e^{-i\theta_0}e^{-i\omega v}\psi^{in}_{\omega l} - e^{i\theta_0}e^{-i\omega u}\psi^{out}_{\omega l} \right).
\end{align}

Next, we consider the behaviour of these physical modes inside the black hole. Rewriting in terms of $U,V$, we get
\begin{align}
w_{\omega lm}&=(4\pi\omega)^{-1/2}r^{-1}Y_{lm}(\theta,\phi)\nonumber\\
&(-i)\left(e^{-i\theta_0}V^{-\frac{i\omega}{2\pi T_H}}\psi^{in}_{\omega l} -
 e^{i\theta_0}(-U)^{\frac{i\omega}{2\pi T_H}}\psi^{out}_{\omega l} \right).
\end{align}
We then analytically continue to the parallel exterior of the black
hole, crossing the singularities at $UV=0$ by analytic continuation in
the lower half-plane in both $U$ and~$V$. Note that the part involving
$\psi^{in}$ is regular across $U=0$ and the part involving $\psi^{out}$
is regular across $V=0$. The rest of the derivation follows quite
similarly to the Schwarzschild case \cite{PhysRevD.15.2088}: we compute
the Bogoliubov coefficients of the physical mode relative to the Kruskal
modes and find
\begin{align}
\left\langle \Psi \right|a^{\dagger}_{\omega lm}a_{\omega lm}\left|\Psi\right\rangle&=\frac{1}{e^{\omega/T_H}-1},\\
\left\langle \Psi \right|a_{\omega lm}a^{\dagger}_{\omega lm}\left|\Psi\right\rangle&=\frac{1}{1-e^{-\omega/T_H}},
\end{align}
where we write the annihilator of the usual physical mode as $a_{\omega lm}$, our state $\ket{\Psi}=\ket{0_K}$ is the Hartle-Hawking vacuum (i.e. the Kruskal vacuum), and all other operator combinations vanish. 
Note that this is a Bose-Einstein distribution, as expected \cite{NavarroSalas}. 

Using the field operator in \eqref{phix} and these operator relations, we arrive at our final result,
\begin{align}
W(x,x')&=\left\langle\Psi\right|\phi(x)\phi(x')\left|\Psi\right\rangle \nonumber\\
&=\sum_{l=0}^{\infty}\sum_{m=-l}^l\int_{0}^{\infty}d\omega\nonumber\\
&\left[\frac{w_{\omega l m}(x)w^*_{\omega l m}(x')}{1-e^{-\omega/T_H}}
+ \frac{w^*_{\omega l m}(x)w_{\omega l m}(x')}{e^{\omega/T_H}-1}\right]\nonumber\\
&=\sum_{l=0}^{\infty}\sum_{m=-l}^l\int_{0}^{\infty}\frac{d\omega}{2\sinh(\omega/2T_H)}\nonumber\\
&\left[e^{\omega/2T_H}w_{\omega l m}(x)w^*_{\omega l m}(x')\right.\nonumber\\
&\left.+ e^{-\omega/2T_H}w^*_{\omega l m}(x)w_{\omega l m}(x')\right].
\end{align}

\bibliography{sads_stationary}
\end{document}